\documentclass[11pt, letter]{article}
\usepackage{jheppub}
\usepackage[english]{babel}
\usepackage{blindtext}\blindmathtrue
\usepackage{avant} 
\usepackage[dvipsnames]{xcolor}
\usepackage{amsmath,epsf,amssymb,mathtools,latexsym,amsthm,setspace,array,pifont,hyperref,amsfonts,dsfont,cancel,braket,parskip}
\usepackage[none]{hyphenat} 
\usepackage{graphicx}
\usepackage{float}
\usepackage{tikz}
\usepackage{tikz-feynman, contour}
\tikzfeynmanset{compat=1.1.0}
\usetikzlibrary{shapes,arrows,positioning,automata,backgrounds,calc,er,patterns}
\makeatletter
\newlength\CoolS@sizex
\newlength\CoolS@sizey
\newcommand*\CoolS@inner{%
\begin{tikzpicture}[baseline=0.04\CoolS@sizey]%
\foreach \x in {0, 1, ..., 5} \foreach \y in {0, 1, ..., 10}
\coordinate (c\x\y) at (\x *0.12*\CoolS@sizex, \y *0.107*\CoolS@sizey);
\draw [line width=\Cool@stroke] (c28)--(c26)--(c44)--(c42)--(c20)--(c02)--(c04)--(c15);
\draw [line width=\Cool@stroke] (c22)--(c24)--(c06)--(c08)--(c210)--(c48)--(c46)--(c35);
\end{tikzpicture}}
\usepackage{todonotes}
\graphicspath{{Pictures/}} 

\newcommand{\nn}{\nonumber}
\newcommand{\sd}{\mathrm{d}}
\newcommand{\pd}{\partial}

\newcommand{\dA}{\dot{\alpha}}
\newcommand{\dB}{\dot{\beta}}

\newcommand{\dD}{\dot{\delta}}
\newcommand{\A}{\alpha}
\newcommand{\B}{\beta}
\newcommand{\G}{\gamma}
\newcommand{\E}{\epsilon}

\newcommand{\D}{\delta}
\newcommand{\R}{\mathbb{R}}
\newcommand{\C}{\mathbb{C}}

\newcommand{\cl}[1]{\mathcal{#1}}

\renewcommand{\Re}{\operatorname{Re}}

\def\prd{\ref@{Phys.~Rev.~D}}        

\renewcommand{\[}{\begin{equation}\begin{aligned}}
\renewcommand{\]}{\end{aligned}\end{equation}}

\usepackage{tcolorbox}
\definecolor{airforceblue}{rgb}{0.36, 0.54, 0.66}
\definecolor{azure}{rgb}{0.0, 0.5, 1.0}
\newtcolorbox{tdbox}{colback=airforceblue!40!white,colframe=azure!90!black} 
\newcommand{\td}[1]{
	\if\notesOn1
	\begin{tdbox}
		#1
	\end{tdbox}
	\fi
}
\hypersetup{
	colorlinks = true,
	linkcolor = Mahogany,
	anchorcolor = Mahogany,
	citecolor = Mahogany,
	filecolor = Mahogany,
	urlcolor = Mahogany
}

\def\notesOn{0}
\tikzset{
	graviton/.style={
		double,
		decoration={snake, aspect=0.75, mirror, segment length=1.5mm},
		decorate
	}
}

\tikzfeynmanset{
	bigblob/.style={
		shape=circle,
		draw=blue,
		fill=red}
}

\title{Scattering Amplitudes and The Cotton Double Copy}
\author[1]{William~T.~Emond}
\author[2,3]{and Nathan Moynihan}
\affiliation[1]{CEICO, Institute of Physics of the Czech Academy of Sciences, Na Slovance 2, 182 21 Praha 8, Czech Republic}
\affiliation[2]{Higgs Centre for Theoretical Physics, School of Physics and Astronomy, The University of Edinburgh, EH9 3FD, Scotland}
\affiliation[3]{School of Mathematics \& Hamilton Mathematics Institute,\\Trinity College Dublin, College Green, Dublin 2, Ireland}
\emailAdd{william.emond@fzu.cz}
\emailAdd{nathantmoynihan@gmail.com}
\abstract{	
	We construct classical curvature spinors in topologically massive gauge theory and topologically massive gravity, expressed in terms of massive three-particle amplitudes. We show that when the amplitudes double copy, the curvature spinors satisfy the \textit{Cotton double copy}, the three-dimensional cousin of the Weyl double copy. Furthermore, we show that under certain circumstances the Cotton double copy can be derived via a dimensional reduction of the Weyl double copy.
}

\begin{document}
\maketitle
\section{Introduction}
Recent developments in scattering amplitudes have led to some intriguing formulations of purely classical physics directly in terms of the on-shell scattering amplitudes \cite{Kosower:2018adc,Maybee:2019jus,Monteiro:2020plf,delaCruz:2020bbn,Cristofoli:2021jas,Britto:2021pud}, including in theories in lower dimensions \cite{Burger:2021wss,Moynihan:2020ejh,Gonzalez:2021bes,Gonzalez:2021ztm,Hang:2021oso}. The double copy, which relates gauge and gravitational scattering amplitudes \cite{Bern:2008qj,Bern:2010yg,Johnson:2020pny,Momeni:2020vvr,Momeni:2020hmc,Moynihan:2020ejh,Bargheer:2012gv,Gonzalez:2021bes,Gonzalez:2022mpa,Chiodaroli:2022ssi,Berman:2022bgy,Moynihan:2020gxj,Chacon:2021lox,Menezes:2021dyp,Brandhuber:2021bsf,Li:2021yfk,Frost:2021qju,Sivaramakrishnan:2021srm,Ahmadiniaz:2021ayd,Witzany:2021dru,Emond:2021lfy,Brandhuber:2021eyq,Borsten:2021rmh,Bastianelli:2021rbt,Chi:2021mio,Hang:2021fmp,Brandhuber:2021kpo,Carrasco:2020ywq,Emond:2020lwi} also has several interesting classical formulations that relate solutions \cite{Monteiro:2014cda,Luna:2015paa,Luna:2016hge,Monteiro:2021ztt,CarrilloGonzalez:2019gof,White:2020sfn,Chacon:2020fmr,Chacon:2021hfe,Cheung:2021zvb,Moynihan:2021rwh,Ben-Shahar:2021zww,Cho:2021nim,Gonzo:2021drq,Chacon:2021hfe,Adamo:2021dfg,Lescano:2022nhp,Escudero:2022zdz,Shi:2021qsb,Diaz-Jaramillo:2021wtl,Alawadhi:2021uie,Angus:2021zhy,Alkac:2021seh,Farnsworth:2021wvs,Almeida:2020mrg,Easson:2020esh,Guevara:2021yud} including the Weyl double copy \cite{Luna:2018dpt,Alawadhi:2019urr,Godazgar:2020zbv,Godazgar:2021iae,Easson:2021asd,Alawadhi:2020jrv,Chacon:2021wbr}, which we will explore here.

The Weyl double copy is formulated in terms of complex spinors built out of the Faraday tensor and the Weyl tensor, and asserts that the Weyl spinor $\psi_{\A\B\G\D}$ is related to the Maxwell spinor $\varphi_{\A\B}$ via the relation \cite{Luna:2018dpt}
\[
\psi_{\A\B\G\D} = \frac{\varphi_{(\A\B}\varphi_{\G\D)}}{S},
\]
where $S(x)$ is a scalar field. In Ref. \cite{Monteiro:2020plf}, it was shown that these classical quantities are related directly to on-shell three-particle amplitudes, which are themselves avatars of the Newman-Penrose scalars. 

In four dimensional Minkowski space, three-particle scattering amplitudes with real momenta are typically never on-shell. There are two choices that are usually made to define on-shell three-particle amplitudes: opting for complex momentum or choosing to work in $(2,2)$ signature \cite{Monteiro:2020plf,Monteiro:2021ztt,Crawley:2021auj}. Working with analytically continued momentum is often favoured, for example when the goal is on-shell constructibility \cite{Britto:2004ap,Britto:2005fq,Carballo-Rubio:2018bmu,Schuster:2008nh,Cheung:2015cba,Cheung:2015ota,Elvang:2018dco,Rodina:2018pcb}, however this is not always an ideal choice, e.g. the contour integration over complex momentum might be difficult when compared with a real integral in split signature. Conceptually, working in split signature also has certain useful virtues: the decomposition of a momentum gives two real, distinct spinors of opposite chirality and polarization vectors are real. Of importance to us will be the fact that classical solutions in gauge theory and gravity have an interpretation in terms of on-shell three-particle amplitudes \cite{Monteiro:2020plf}.

One important thing to note is that for real momentum vectors, only one of the components of the momentum becomes imaginary in spinor space, which is why working in $(2,2)$ makes everything real: you simply analytically continue that component (or the other three). Working in spinor-helicity, it is straightforward to dimensionally reduce four-dimensional amplitudes to three-dimensions, where massless gauge bosons become topologically massive under a reduction. If we choose the direction of spacetime that would otherwise be analytically continued in four dimensions, we end up with bispinors in three-dimensions whose dynamical components are real (in the representation $SL(2,\R)$) and three-particle amplitudes can in principle have support on-shell. 

In \cite{Burger:2021wss}, the present authors derived various classical solutions relating to anyons using spinor helicity variables in the (complex) $SU(1,1)$ representation. However, each representation is related by a Cayley transformation, and so these solutions can readily be recycled into $SL(2,\R)$ form.

In this paper, we wish to study the analogue of the Weyl double copy in three spacetime dimensions. This isn't obviously a sensible idea, since there are several issues: the Weyl tensor is identically zero in three dimensions, in any signature, and there is no possible analogue of split signature. However, since three-particle scattering amplitudes in topologically massive theories do enjoy a double copy \cite{Moynihan:2020ejh,Burger:2021wss,Gonzalez:2021bes,Gonzalez:2021ztm} and furthermore that they can be obtained sensibly from a dimensional reduction from four dimensions \cite{Moynihan:2020ejh,Burger:2021wss}, it is an intriguing question as to what the 3D analogue of the Weyl double copy is?

\section{Motivation: Topological Mass via Dimensional Reduction}\label{motivation}
Equations of motion for free fields can be greatly simplified by using the relationship between tensors with $s$ indices and spinors with $2s$ indices. In four dimensions, this corresponds to representing every vector index with a pair of $SL(2,\C)$ indices, one in the $(\frac12,0)$ representation and the other in $(0,\frac12)$, typically achieved by contracting the vector index with an Infeld–Van der Waerden symbol $\sigma^N_{\A\dB}$, where $N = 0,1,2,3$ is a $D = 4$ Lorentz index and $\A$ and $\dB$ are spinor indices, manipulated using $\epsilon_{\A\B}$ and $\epsilon_{\dA\dB}$ respectively. Inner products between Lorentz tensors can easily be exchanged for spinor contractions, and vice versa, using the relationships
\begin{equation}\label{key}
	\eta^{MN} = -\frac12\sigma^{M}_{\A\dB}\sigma^{N \A\dB},~~~~~\sigma^M_{\A\dB}\sigma_{M \G\dD} = -2\epsilon_{\A\G}\epsilon_{\dB\dD}.
\end{equation}
Using this, we can recast the Maxwell equations and Bianchi identities into the form
\begin{align}\label{key}
	\pd_N F^{MN} = 0 ~~~&\longleftrightarrow~~~\pd_{\dB}^{\A}\varphi_{\A\B} + \pd_{\B}^{\dA}\bar{\varphi}_{\dA\dB} = 0,\\
	\pd_{[M}F_{NR]} = 0 ~~~&\longleftrightarrow~~~\pd_{\dB}^{\A}\varphi_{\A\B} - \pd_{\B}^{\dA}\bar{\varphi}_{\dA\dB} = 0,
\end{align}
where $\varphi_{\A\B}$ is the self-dual Maxwell spinor (and $\bar{\varphi}_{\dA\dB}$ its complex conjugate, the anti-self-dual spinor) given by
\begin{equation}\label{key}
	\varphi_{\A\B} = \varphi_{(\A\B)} = \frac12\epsilon^{\dA\dB}F_{MN}\sigma^M_{\A\dA}\sigma^N_{\B\dB}~.
\end{equation}
By combining both equations for $\varphi_{\A\B}$, we can derive a single equation of motion for the self-dual sector
\begin{equation}\label{key}
	\pd_{\dB}^{\A}\varphi_{\A\B} = 0.
\end{equation} 
In $2+1$ dimensions the situation is a bit different, and the Lorentz group has several interesting isomorphisms $$SO(1,2) \sim SL(2,\R) \sim SU(1,1) \sim SP(2,\R).$$
We will choose to work in the $SL(2,\R)$ representation (with no distinction between left and right indices), meaning the components of a momentum bi-spinor $p_{\A\B}$ are all real assuming the momentum vector has real components. We can then use the (2+1)-dimensional relationship between Lorentz and $SL(2,\R)$ inner products
\begin{equation}\label{key}
	\eta^{\mu\nu} = -\frac12\sigma^{\mu}_{\A\B}\sigma^{\nu \A\B},~~~~~\sigma^\mu_{\A\B}\sigma_{\mu \G\D} = -2\epsilon_{\A\G}\epsilon_{\B\D},
\end{equation} 
where $\mu = 0,1,2$ and the Infeld–Van der Waerden symbols $\sigma^\mu_{\A\B}$ are completely real in the $SL(2,\R)$ basis given in the appendix.

The equations of motion for topologically massive gauge theory also have an $SL(2,\R)$ representation, as does the 3D Bianchi identity\footnote{To see this, note that the Bianchi identity in $2+1$ dimensions can be written as
\begin{equation}\label{key}
	\pd_\mu\tilde{F}^\mu \propto  \epsilon^{\mu\nu\rho}\pd_\mu F_{\nu\rho} \propto \pd^{\A\B}\varphi_{\A\B} = 0,~~~~~ \implies ~~~~~ \epsilon^{\A\B}\left(\pd_{\B}^{\G}\varphi_{\G\A} - \pd_{\A}^{\G}\varphi_{\G\B}\right) = 0.
\end{equation}
}
\begin{align}\label{key}
	\pd_\nu F^{\mu\nu} + m\tilde{F}^\mu = 0 ~~~&\longleftrightarrow~~~\pd_{\B}^{\G}\Phi_{\G\A} + \pd_{\A}^{\G}\Phi_{\G\B} - 2m\Phi_{\A\B} = 0,\\
	\pd_\mu \tilde{F}^\mu = 0 ~~~&\longleftrightarrow~~~\pd_{\B}^{\G}\Phi_{\G\A} - \pd_{\A}^{\G}\Phi_{\G\B} = 0,
\end{align}  
where $\Phi_{\A\B}$ is completely real and the dual to the Faraday tensor is given by
\begin{equation}\label{key}
\tilde{F}^\mu \equiv	\frac{1}{2}\epsilon^{\mu\rho\sigma}F_{\rho\sigma}.
\end{equation}
The Bianchi identity simplifies the equations of motion to give a simple equation of motion
\begin{equation}\label{key}
	\pd_{\A}^{\G}\Phi_{\G\B} -m\Phi_{\A\B} = 0.
\end{equation}
We note that since $\Phi_{\A\B}$ is real, it does not change when considering a helicity flip (corresponding to complex conjugation), unlike in four dimensions. As in the standard case for the tensorial equations of motion, where $F^{\mu\nu}$ is real, the helicity is determined explicitly by the sign of the mass.
\subsection{Spinor Dimensional Reduction}

The convenience of working in the $SL(2,\R)$ representation is that it can be considered as the real slice of four dimensional spinor space, at least for vectors and antisymmetric two-tensors with real components. For a generic spacetime vector $\ell^\mu$ (with real components), we can perform a dimensional reduction along a given spacetime direction by taking a real slice via\footnote{A simple example is the position vector $x^M = (t,x,y,z)$ 
\begin{equation}\label{key}
	\Re\begin{pmatrix}
		-t +z & x+iy \\
		x-iy & -t-z
	\end{pmatrix} = \begin{pmatrix}
	-t +z & x \\
	x & -t-z
\end{pmatrix}.
\end{equation}
}, i.e.
\begin{equation}\label{key}
	\Re\left(\ell_{4D}^M\sigma_{M}^{\A\dB}\right) = \ell_{3D}^{\A\dB}.
\end{equation} 
However, we still have the chiral `dotted' index to deal with, which ought not to be present in a 3D representation. We can introduce an operator
\begin{equation}\label{dimRedOp}
	\chi^{~\dA}_{\A} = \chi^M(\gamma_{M})^{~\dA}_{\A} = (\gamma_{2})^{~\dA}_{\A},
\end{equation}
which projects out the specified imaginary direction such that a symmetrization automatically performs the dimensional reduction, e.g.
\begin{equation}\label{key}
	\ell_{\A\dA}\chi^{\dA}_\B = \ell_{\A\B} + i\ell_2\epsilon_{\A\B}~~~\implies~~~\ell_{(\A\dA}\chi^{\dA}_{\B)} = \ell_{\A\B}.
\end{equation}

If we want to do a dimensional reduction from a massless vector to a massive, we take the direction of spacetime to be constant rather than zero, which gives
\begin{equation}\label{key}
	\det{p_{\A\dA}} = -p_0^2 + p_1^2 + p_2^2 + p_3^2 = 0 ~~~~\longleftrightarrow~~~~ -p_0^2 + p_1^2 + p_2^2 = -m^2.
\end{equation}
In position space, this is simply 
 \begin{equation}\label{key}
 	\chi^{~\dA}_{\B}\pd_{\dA\A} = \pd_{\A\B} \pm m\epsilon_{\A\B},
 \end{equation} 
where the sign of $m$ corresponds to either an outgoing or incoming plane-wave (and again determines the helicity).
 
We can use this to derive the spinor form of the topologically massive equations of motion directly from four dimensional self-dual Maxwell theory, since we can write
\begin{equation}\label{key}
	\Re\left(\chi^{~\dB}_{\A}\pd_{\dB}^{\G}\varphi_{\G\B}\right) = \pd_{\A}^{\G}\Phi_{\G\B} - m\Phi_{\A\B} = 0.
\end{equation}
This is only straightforwardly possible due to the fact that we are examining the spinor form of a completely antisymmetric tensor with real components. This tells us that all of the imaginary components of the resulting spinor will lie along one direction of spacetime - if it had been a symmetric tensor, or a tensor with imaginary components, this would not have necessarily been the case. This makes examining the gravitational case more complicated, since the Weyl spinor comes from a tensor which is symmetric in some of its indices. 

Symmetric curvature spinors can, in general, always be decomposed in terms of rank-1 spinors in an effort to classify exact solutions of the field equations \cite{Penrose:1960eq,Witten:1959zza}, including in topologically massive gravity \cite{Barrow_1986,Milson:2012ry,Chow:2009km}. The Weyl curvature spinor can be decomposed uniquely into a symmetrized product of one-index spinors \cite{Penrose:1960eq}
\begin{equation}\label{key}
	\Psi_{\A\B\G\D} = \A_{(\A}\B_\B\G_\G\D_{\D)}.
\end{equation}
The character of the rank one spinors, namely how they specify principal null directions, determines the possible Petrov class of the Weyl tensor. Since these spinors point out directions in spacetime, they ought to have a natural dimensional reduction of the form discussed. However, as discussed in the introduction the Weyl tensor is identically zero in less than four dimensions making this question less trivial. In three dimensions, the role of the Weyl tensor is typically taken up by the Cotton tensor \cite{Garcia:2003bw,Sousa:2007ax}, which shares many of the desirable properties (e.g. its vanishing indicates conformal flatness), so it is natural to expect that solutions in topologically massive gravity might be classified using the Cotton spinor. To examine this problem, we will express the curvature spinors explicitly in terms of scattering amplitudes and momentum spinors, which are well behaved under dimensional reduction \cite{Moynihan:2020ejh,Burger:2021wss}.


\section{Curvature Spinors from Three-Particle Amplitudes}
Curvature spinors can decomposed into rank one spinors, each of which is associated to some null vector in spacetime. Null vectors in four dimensions give rise to bi-spinors with vanishing determinant and can therefore always be decomposed into rank one spinors. It is useful to construct scattering amplitudes using spinor-helicity variables -- the spinors that correspond to a decomposition of a null momentum in four dimensions. In three dimensions, we can decompose massive momentum bispinors via
\[
	p_{\A\B} = N\lambda_{(\A}\bar{\lambda}_{\B)},
\]
where $N$ is a dimensionless normalisation to be determined.
The on-shell condition, in the signature $(-,+,+)$, demands that
\[
-2p^\mu p_\mu = p^{\A\B}p_{\A\B} = 2m^2 = -2N^2\braket{\lambda\bar{\lambda}}^2.
\]
If we demand that the momentum bispinor is real and that it satisfies on the on-shell condition, then a suitible choice that gives rise to the momentum in the basis \eqref{paulibasis} is
\[
\lambda_\A = \frac{1}{\sqrt{p_1-p_0}}\begin{pmatrix}
p_2-im \\ p_1-p_0
\end{pmatrix},~~~~~\bar{\lambda}_\A = \frac{1}{\sqrt{p_1-p_0}}\begin{pmatrix}
p_2+im \\ p_1-p_0
\end{pmatrix}.
\]
With $\braket{\lambda\bar{\lambda}} = 2im$, the normalisation $N = \frac{1}{2}$ satisfies the on-shell condition. We find a momentum bispinor of the form
\[
p_{\A\B} = \frac{1}{2}\lambda_{(\A}\bar{\lambda}_{\B)} = \begin{pmatrix}
-p_0-p_1	&  p_2\\
p_2	& -p_0 + p_1
\end{pmatrix} = p_\mu\sigma^\mu_{\A\B}.
\]
The polarization vectors for these spinors are given by \cite{Moynihan:2020ejh}
\begin{equation}\label{polvectors}
	\epsilon_-^\mu(q) = \frac{\braket{q|\gamma^\mu|q}}{2m},\quad \epsilon_+^\mu(q) = \frac{\braket{\bar{q}|\gamma^\mu|\bar{q}}}{2m}\quad \Longleftrightarrow\quad\epsilon_{- \A\B}(q) = \frac{|q\rangle_\A|q\rangle_\B}{m}	,\quad \epsilon_{+ \A\B}(q) = \frac{|\bar{q}\rangle_\A|\bar{q}\rangle_\B}{m} \;.
\end{equation}
They complex conjugates of one another, with conjugation exchanging incoming for outgoing.

We could of course choose to ensure that our spinors are entirely real by analytically continuing the mass to $m\rightarrow -im$. This either causes the momentum bispinor to become pure imaginary (since we now require $N = i/2$) or we keep $N = 1/2$ and the bispinor real and the on-shell condition is violated, giving $p^2 = m^2 \neq -m^2$. In the following subsections we will opt for the spinors to be complex -- including their momentum components -- and comment on the other cases later.

\subsection{Electromagnetic Curvature Spinor}
Matter fields coupled to topologically massive gauge bosons are anyons due to the presence of the Chern-Simons term \cite{Jackiw:1990ka}. We can do the usual mode expansion of a topologically massive gauge field into incoming and outgoing modes, 
\begin{equation}\label{key}
	A_\mu(x) = \int \sd\Phi(q)\left[a(q)\epsilon^\mu(q)e^{-iq\cdot x} + a^\dagger(q)\epsilon^{*\mu}(q)e^{iq\cdot x}\right],
\end{equation}
where the measure is
\begin{equation}\label{key}
	\sd\Phi(q) = \hat{\sd}^3q\hat{\delta}(q^2 + m^2)\Theta(q^0),~~~~~\hat{\sd}^nq \equiv \frac{\sd^nq}{(2\pi)^n},~~~~~\hat{\delta}^n(q) \equiv (2\pi)^n\delta^n(q).
\end{equation}
We note that this expansion is the same for either value of the spin of the gauge field (which corresponds to the sign of the mass in the Lagrangian - no sum over helicities), with the difference encapsulated by the polarization vector $\epsilon^\mu$.

In 2+1 dimensions, the dual field strength vector is equivalent to the field strength tensor, and so we will use this to derive the corresponding Maxwell spinor. Thankfully, the dual field strength can be expanded into modes just as the gauge field can, and in fact by using the equations of motion
\begin{equation}\label{key}
	\tilde{F}^\mu = -\frac{1}{m}\pd^2 A^\mu(x) + \frac{1}{m}\pd^\mu\pd_\nu A^\nu(x), 
\end{equation}
we can write on-shell
\[
\tilde{F}^\mu(x) &= \int \sd\Phi(q) \frac{q^2}{m}\left[a(q)\epsilon^\mu(q)e^{-iq\cdot x} + a^\dagger(q)\epsilon^{*\mu}(q)e^{iq\cdot x}\right]\\
&= -mA^\mu(x).
\]
In four dimensions, the expectation value of $F^{\mu\nu}$ can be expressed in terms of a three-particle amplitude \cite{Monteiro:2020plf}
\[
\left\langle F^{\mu \nu}(x)\right\rangle=\frac{1}{M} \operatorname{Re} \sum_{\eta} \int \mathrm{d} \Phi(\bar{k}) \hat{\delta}(u \cdot \bar{k}) \mathcal{A}_{-\eta}^{(3)}(\bar{k}) \bar{k}^{[\mu} \varepsilon_{\eta}^{\nu]} e^{-i \bar{k} \cdot x},
\]
where $\eta$ is the helicity and $\bar{k}^\mu$ a wavevector. We can analogously treat the dual of $F^{\mu\nu}$ in the same way, but in three-dimensions (see also \cite{Burger:2021wss}), taking it to be
%
\begin{equation}\label{key}
	\braket{\tilde{F}^\mu(x)}_{k} = \frac{m}{M}\Re\int \sd\Phi(\bar{q})\delta(u\cdot \bar{q})\cl{A}^{(3)}_{-k}(\bar{q})\epsilon_k^\mu(\bar{q})e^{-i\bar{q}\cdot x},
\end{equation}
where we have explicitly labelled the intrinsic helicity $k = \pm1$, where incoming (outgoing) particles have helicity $+k$ ($-k$). We also note the overall factor of $m$, which comes from the relationship between $\tilde{F}^\mu$ and $A^\mu$.

For the remainder of the paper we will take $k = -1$ and we will take $q^\mu$ to be a classical wavevector where contextually appropriate (with no bar notation to avoid confusion with barred spinors). We can then express the field strength as\footnote{We also note that we could have arrived at this result directly by dualising the usual Faraday tensor, i.e.
	\[
	\braket{\tilde{F^\mu}} &= \frac12\epsilon^{\mu\nu\rho}\braket{F_{\nu\rho}} = \frac{1}{4m}\epsilon^{\mu\nu\rho}\Re\int \sd\Phi(\bar{q})\delta(p_1\cdot q)\cl{A}^{(3)}_{+}(q)(q_\nu \braket{q|\gamma_\rho|q}-q_\mu \braket{q|\gamma_\nu|q})\\
	&=  \frac12\Re\int \sd\Phi(\bar{q})\delta(p_1\cdot q)\cl{A}^{(3)}_{+}(q)\bra{q}\gamma^\mu\ket{q}e^{-iq\cdot x}
	\]
	where we have used
	\[
	\epsilon^{\mu\nu\rho}q_\nu\braket{q|\gamma_\rho|q} = q_\nu\braket{q|\gamma^\mu\gamma^\nu|q} = \braket{q|\gamma^\mu q|q} = 2im\braket{q|\gamma^\mu|q}.
	\]}
\begin{equation}\label{eq:pert ecp value gauge}
	\braket{\tilde{F}^\mu(x)} = \frac{1}{2M}\Re\int \sd\Phi(q)\delta(u\cdot q)\cl{A}^{(3)}_{+}(q)\braket{q|\gamma^\mu|q}e^{-iq\cdot x}.
\end{equation}

Acting with $\sigma_{\A\B}$ then gives the Maxwell spinor in terms of the on-shell three-particle amplitude
\begin{equation}\label{key}
	\Phi_{\A\B} = \braket{\tilde{F}^\mu\sigma_{\mu \A\B}} = \frac{1}{M}\Re\int \sd\Phi(\bar{q})\delta(u\cdot q)e^{-iq\cdot x}\ket{q}_\A\ket{q}_\B \cl{A}^{(3)}_{+}(q).
\end{equation}
As an example, which will shall explicitly verify later, we can consider a static scalar coupled to a topologically massive gauge boson --- an anyon --- whose three-particle amplitudes are given by \cite{Moynihan:2020ejh,Burger:2021wss}
\begin{equation}\label{key}
	\cl{A}_3[1,1',q^\pm] = eMx^\pm, ~~~~~x \equiv \frac{\braket{\bar{q}|u_1|\bar{q}}}{m},~~~~~x^{-1} \equiv \frac{\braket{q|u_1|q}}{m},
\end{equation}
where $q^2 = -m^2$, $p_1^2 = p_1'^2 = -M^2$. Plugging these in then gives the curvature spinor for a static scalar anyon
\[\label{anyonspinor}
	\Phi_{\A\B} = \frac{e}{m}\Re\int \sd\Phi(\bar{q})\delta(u\cdot q)e^{-iq\cdot x}\ket{q}_\A\ket{q}_\B \braket{\bar{q}|u_1|\bar{q}}.
\]

\subsection{Gravitational Curvature Spinor}
The linearized graviton field operator for a single helicity is given by 
\begin{equation}\label{eq:graviton}
	h^{\mu\nu} = 2\,\text{Re}\int\sd\Phi(q)\,a(q)\epsilon^{\mu}(q)\epsilon^{\nu}(q)\,e^{-iq\cdot x} \;,
\end{equation}
where we have expressed the graviton polarisation tensor as an outer product of polarisation vectors $\epsilon^{\mu}(q)$.

The linearized Cotton tensor $C_{\mu\nu}$ is given by
\begin{equation}
	C^{\mu\nu}  = \frac{\kappa}{2}\,\pd_\lambda\big[\pd^\rho\varepsilon^{\lambda\sigma(\mu}\pd^{\nu)}h_{\rho\sigma} - \pd^2\varepsilon^{\lambda\sigma(\mu}h^{\nu)}_{\;\;\sigma}\big] \;,
\end{equation}
which in the de Donder gauge ($\pd_\nu h^{\mu\nu}=\frac{1}{2}\pd^\mu h$), reduces to 
\begin{equation}
	C^{\mu\nu}  = -\frac{\kappa}{4}\,\pd_\lambda \pd^2\varepsilon^{\lambda\sigma(\mu}h^{\nu)}_{\;\;\;\sigma} \;.
\end{equation}
Inserting \eqref{eq:graviton} into this expression gives us the Cotton tensor operator for a given fixed helicity $k$
\begin{align}
	C^{\mu\nu}_k =& -\frac{i\kappa}{2}\,\text{Re} \int\sd\Phi(q)\,a_{j}(q)\,q^2q_\lambda\,\varepsilon^{\lambda\sigma(\mu}\epsilon_k^{\nu)}(q)\epsilon_{k\sigma}(q)\,e^{-iq\cdot x} \nn\\ =& \, \frac{i\kappa m^2}{2}\,\text{Re} \int\sd\Phi(q)\,a_{k}(q)\,q_\lambda\,\varepsilon^{\lambda\sigma(\mu}\epsilon_k^{\nu)}(q)\epsilon_{k\sigma}(q)\,e^{-iq\cdot x}\;,
\end{align}
where we have used that $q^2=-m^2$, with $m$ is the topological mass of the graviton.
To compute the corresponding expectation value of the Cotton tensor, we can do so in direct analogy to the gauge theory case for $C^{\mu\nu}(x)$. 
In doing so, one arrives at the following expression for the expectation value of the Cotton tensor 
\begin{equation}
	\langle C^{\mu\nu}\rangle_k = \,-\frac{\kappa m^2}{4M}\,\text{Re}\int\sd\Phi(q)\,\hat{\delta}(u\cdot q)\,\mathcal{M}_{-k}^{(3)}(q)\,q_\lambda\,\varepsilon^{\lambda\sigma(\mu}\epsilon_k^{\nu)}(q)\epsilon_{k\sigma}(q)\,e^{-iq\cdot x} \;.
\end{equation}
Using the polarization vectors defined in \eqref{polvectors}, along with the identities $\varepsilon^{\mu\nu\rho}\sigma_{\rho\,\A\B} = \eta^{\mu\nu}\epsilon_{\A\B} + (\sigma^\mu\sigma^\nu)_{\A\B}$ and $\sigma^{\mu}_{\A\B}\,\sigma_{\mu\,\G\D} = 2\varepsilon_{\A\D}\varepsilon_{\B\G} + \varepsilon_{\A\B}\varepsilon_{\G\D}=-\varepsilon_{\A(\D}\varepsilon_{\G)\B}$, we find
\begin{equation}
	q_\lambda\,\varepsilon^{\lambda\sigma(\mu}\epsilon^{\nu)}(q)\epsilon_{\sigma}(q)\,\sigma_{\mu\,\A\B}\,\sigma_{\nu\,\G\D} = q_{\A\E}\,\epsilon_{\;\B}^{\E}\,\epsilon_{\G\D} +  q_{\G\E}\,\epsilon_{\D}^{\E}\,\epsilon_{\A\B} \;,
\end{equation}
such that for $k = -1$
\begin{align}
	\mathcal{M}_{+}^{(3)}(q)\,q_\lambda\,\varepsilon^{\lambda\sigma(\mu}\epsilon_-^{\nu)}(q)\epsilon_{-\sigma}(q)\,\sigma_{\mu\,\A\B}\,\sigma_{\nu\,\G\D} = \frac{2}{m}&\ket{q}_\A\ket{q}_\B\ket{q}_\G\ket{q}_\D\,i\mathcal{M}_{+}^{(3)}(q),
\end{align}
with the $k = +1$ being given analogously in terms of barred spinors.

The expectation value of the Cotton spinor is then simply
\begin{align}
	\langle\Psi_{\A\B\G\D}\rangle =& \ \langle C^{\mu\nu}\sigma_{\mu\,\A\B}\,\sigma_{\nu\,\G\D}\rangle \nn\\[0.5em] =& \ -\frac{\kappa}{2} \frac{m}{M}\,\text{Re}\,\int\sd\Phi(q)\,\hat{\delta}(u\cdot q)\Big[\ket{q}_\A\ket{q}_\B\ket{q}_\G\ket{q}_\D\,i\mathcal{M}_{+}^{(3)}(q)\Big]\,e^{-iq\cdot x}\;.
\end{align}
We can again construct the scattering amplitude for a static gravitational anyon directly \cite{Moynihan:2020ejh,Burger:2021wss}
\[
\cl{M}_+^{(3)}(q) = \kappa M^2 x^2 = \kappa M^2\frac{\braket{\bar{q}|u|\bar{q}}^2}{m^2},
\]
which gives the Cotton spinor for the gravitational anyon
\[\label{cotton_anyon}
\Psi_{\A\B\G\D} = -\frac{\kappa^2}{4} \frac{M}{m}\,\text{Re}\,i\int\sd\Phi(q)\,\hat{\delta}(u\cdot q)\Big[\ket{q}_\A\ket{q}_\B\ket{q}_\G\ket{q}_\D\,\braket{\bar{q}|u|\bar{q}}^2\Big]\,e^{-iq\cdot x}\;.
\]
\subsubsection{Momentum Space Weyl Double Copy}
If we further define the momentum-space version of the Cotton spinor by
\begin{equation}
	\langle\Psi_{\A\B\G\D}\rangle = \kappa\,\text{Re}\int\sd\Phi(q)\,\hat{\delta}(2p\cdot q)\,\Psi_{\A\B\G\D}(q)\,e^{-iq\cdot x} \;,
\end{equation}
then it is immediately clear that
\begin{equation}
	\Psi_{\A\B\G\D}(q) = -im\ket{q}_\A\ket{q}_\B\ket{q}_\G\ket{q}_\D\,\mathcal{M}_{+}^{(3)}(q),
\end{equation}
which is a type N representation in momentum-space.

We can compare this with the gauge theory momentum space curvature spinor given by
\[
\Phi_{\A\B}(q) = \ket{q}_\A\ket{q}_\B\cl{A}_+^{(3)}(q).
\]
Following \cite{Monteiro:2020plf}, we can define the scalar analogue as
\[
S(x) = \Re\int\sd\Phi(q)\,\hat{\delta}(2p\cdot q)S(q)e^{-iq\cdot x} \;,
\]
such that we establish the momentum space Cotton double copy
\[
\Psi_{\A\B\G\D}(q) = \frac{m}{2}\frac{\Phi_{(\A\B}(q)\Phi_{\G\D)}(q)}{S(q)},
\]
which follows from the fact that the three-particle scattering amplitudes double copy \cite{Moynihan:2020ejh,Burger:2021wss}.
\subsubsection{A Position Space Example}
Having established the momentum-space Cotton double copy --- expecting it to hold in position space --- we now motivate this position with a specific example, namely that of a plane-fronted wave in 2+1 dimensions  \cite{Deser:1993wt,Deser:1992nk,Deser:2004wd}. In Kerr-Schild coordinates, the plane-wave metric can be written as \cite{Deser:1992nk}
\[
g_{\mu\nu} = \eta_{\mu\nu} + \kappa\delta(\ell\cdot x)H(x^\mu)\ell_\mu \ell_\nu,
\]
where $\ell_\mu$ is a unit null vector the lies along the lightcone, e.g. $\ell_\mu = \pd_\mu(t-x) = (1,-1,0)$.

In order to find a form for $H(x^\mu)$, we need to solve the equations of motion, which is built from $C_{\mu\nu}$ and $G_{\mu\nu}$.  These are given by
\[
C_{\mu\nu} &= -\frac{\kappa}{2}\delta(\ell\cdot x)\varepsilon^{\lambda\sigma}_{~~\mu} \pd_\lambda\pd^2 H(x)\ell_\sigma\ell_\nu\\
G_{\mu\nu} &= -\frac{\kappa}{2}\delta(\ell\cdot x)\ell_\mu \ell_\nu \pd^2 H(x),
\]
where we note that the derivative of the delta function doesn't play a role since $\ell_\mu$ is a null vector. The delta function constraint, together with the condition that $\ell$ lies along a lightcone, means that $F$ will only ever depend on one direction --- either $x$ or $y$. 
This means we need to solve the vacuum topologically massive gravity field equations
\[
G_{\mu\nu} + \frac{1}{m}C_{\mu\nu} = -\frac{\kappa}{2}\delta(\ell\cdot x)\ell_\mu\ell_\nu\left[\frac{1}{m}H'''(|x|) + H''(|x|)\right] = 0
\]
and we are taking $|x| = \sqrt{x_\mu x^\mu}$. 

Assuming $x>0$, the homogeneous part has the solution
\[
H(|x|) = \kappa e^{-m|x|} + c_1|x| + c_2,
\]
and we reiterate that $|x| = x$ or $|x| = y$ depending on the direction of the wave dictated by $\ell_\mu$. We will take $c_1 = c_2 = 0$ as they are not interesting solutions. 

On the gauge theory side, we need to solve the equation of motion for topologically massive electrodynamics
\[
(\pd^2 - m^2)\tilde{F}^\mu = 0.
\]
Recalling that on-shell we have $\tilde{F}^\mu(x) = -m A_\mu(x)$, in vacuum this is clearly solved by $H(|x|)$ (with $\kappa \rightarrow e$), and so we find a single copy of the form
\[
\tilde{F}^\mu = e\delta(\ell\cdot x)e^{-m|x|}\ell^\mu.
\]
This establishes the Kerr-Schild double copy for plane waves (see also \cite{Gonzalez:2021ztm}) with solution\footnote{We could also include colour factors here, where the scalar field would be a bi-adjoint scalar zeroth copy \cite{Moynihan:2021rwh}.}
\[
h_{\mu\nu} = \kappa\delta(\ell\cdot x)e^{-m\sqrt{x^2}}\ell_\mu\ell_\nu,~~~~~A_\mu = e\delta(\ell\cdot x)e^{-m\sqrt{x^2}}\ell_\mu,~~~~\phi(x) = \lambda\delta(\ell\cdot x)e^{-m\sqrt{x^2}}.
\]
The Maxwell spinor in this case is then given by
\[
\Phi_{\A\B} = -2m\delta(\braket{\ell|x|\ell})e^{-m\sqrt{x^2}}\ket{\ell}_\A\ket{\ell}_\B,
\]  
and the Cotton spinor by
\[
\Psi_{\A\B\G\D} &= -\frac{\kappa}{2}\delta(\ell\cdot x)\varepsilon^{\lambda\sigma}_{~~\mu}\sigma^\mu_{\A\B}\pd_\lambda\pd^2 H(x)\ell_\sigma\ell_{\G\D}\\
&= \kappa^2 m^3\delta(\braket{\ell|x|\ell}) e^{-m\sqrt{x^2}}\hat{x}_\A^{~\epsilon}\ket{\ell}_\epsilon\ket{\ell}_\B\ket{\ell}_\G\ket{\ell}_\D\\
&= \kappa^2 m^3\delta(\braket{\ell|x|\ell}) e^{-m\sqrt{x^2}}\ket{\ell}_\A\ket{\ell}_\B\ket{\ell}_\G\ket{\ell}_\D,
\] 
where we have used the identity in eq. \eqref{epsID} along with $\pd_\lambda H(x) = -m \hat{x}_\lambda H(x)$. On the last line, we have used the fact that the delta function constraint precisely fixes $x\ket{\ell} = |x|\ket{\ell}$. We see immediately that the Cotton double copy holds, since
\[
\Psi_{\A\B\G\D}(x) = \frac{m}{2}\frac{\Phi_{\A\B}(x)\Phi_{\G\D}(x)}{\phi(x)}.
\]
\section{Radiative Solutions and Curvature Spinors}
Having established the relationship between scattering amplitudes in topologically massive gauge theories and curvature spinors, we now need to check from the field theory side that these do indeed give rise to the expected results. Bear in mind that we did not, in fact, use information from any \textit{specific} theory to construct the curvature spinors above: we constructed the three-particle amplitudes for an arbitrary parity-violating massive spin-1 (and spin-2) theory coupled to scalars using little group constraints and dimensional analysis. We will now construct the same quantities for specific solutions from the field theory side.
\subsection{Electromagnetic Curvature Spinor}
Let us consider the (Lorenz gauge) field equations for the topologically massive dual Maxwell field $\tilde{F}^\mu(x)$ in the presence of a static, electrically charged point source $J^\mu(x)$
\begin{equation}
	(\pd^2-m^2)\tilde{F}^\nu(x) = e J^\mu(x)\;,
\end{equation} 
where $J^\mu(x)=(m\eta^{\mu\nu}-\varepsilon^{\mu\nu\rho}\pd_\rho)u_\nu\,\delta^{(2)}(\mathbf{x})$.
We are interested in field solutions that extend to the distant future but not the distant past, so we consider solutions involving the retarded Greens function $G_{\text{ret}}(x-y)$ with support only inside the future lightcone  $x_0 > y_0$, giving a solution
\begin{equation}\label{eq:ret sol}
	\tilde{F}^\mu(x) = -e\int\sd^3y\,G_{\text{ret}}(x-y)\,J^\mu(y) \;.
\end{equation}
The retarded propagator satisfies the equation $(\pd^2-m^2)G_{\text{ret}}(x-y)= -\delta^{(3)}(x-y)$ (with boundary condition $G_{\text{ret}}(x-y)=0$ for $x^0<y^0$), and can therefore be expressed as 
\begin{equation}
	G_{\text{ret}}(x-y) = \lim_{\epsilon\rightarrow 0}\int\hat{d}^3q\frac{e^{iq\cdot (x-y)}}{-(q^0+i\epsilon)^2+\mathbf{q}^2+m^2}\;.
\end{equation}
One can also consider the advanced propagator  $G_{\text{adv}}(x-y)$, which satisfies the same equation, but with the boundary condition $G_{\text{adv}}(x-y)=0$ for $x^0>y^0$:
\begin{equation}
	G_{\text{adv}}(x-y) = \lim_{\epsilon\rightarrow 0}\int\hat{d}^3q\frac{e^{iq\cdot (x-y)}}{-(q^0-i\epsilon)^2+\mathbf{q}^2+m^2}\;.
\end{equation}
The radiative propagator is defined as the difference between these two propagators, i.e.,
\begin{equation}
	G_{\text{rad}}(x-y) = G_{\text{ret}}(x-y) - G_{\text{adv}}(x-y) = i\int\hat{d}^3q \,\text{sgn}(q^0)\hat{\delta}(q^2+m^2)\,e^{iq\cdot (x-y)}\;.
\end{equation}
Note that $G_{\text{rad}}(x-y)$ only has support for on-shell ($q^2=-m^2$) Fourier modes, and as such solves the wave equation $(\pd^2-m^2)G_{\text{rad}}(x-y)=0$.

Given this, we can express eq.~\eqref{eq:ret sol} as 
\begin{equation}
	\tilde{F}^\mu(x) = -e\int\sd^3y\,\big[G_{\text{rad}}(x-y)+G_{\text{adv}}(x-y)\big]\,J^\mu(y) \;.
\end{equation}
Consequently, the radiative contribution to the solution is 
\begin{flalign}\label{eq:rad sol}
	\tilde{F}_{\text{rad}}^\mu(x) =& \  -e\int\sd^3y\,G_{\text{rad}}(x-y)\,J^\mu(y) = -ie\int\hat{\sd}^3q \,\text{sgn}(q^0)\hat{\delta}(q^2+m^2)\,J^\mu(q) \,e^{iq\cdot x} \nn\\[0.5em] =& \ -ie\int\hat{\sd}^3q \,\left(\theta(q^0)-\theta(-q^0)\right)\hat{\delta}(q^2+m^2)\,J^\mu(q) \,e^{iq\cdot x} \nn\\[0.5em] =& \ -ie\int\sd\Phi(q)\,\big[J^\mu(q)\,e^{iq\cdot x} - J^\mu(-q)\,e^{-iq\cdot x}\big]\;.
\end{flalign}
In the case we are considering, the momentum space source term is given by
\begin{equation}
	J^\mu(q)= \delta(q\cdot u)(mu^{\mu}+i\varepsilon^{\mu\nu\rho}u_\nu q_\rho) \;,
\end{equation}
where $u^\mu=(1,0,0)$. Observe that $(J^\mu(q))^\ast = J^\mu(-q)$, and so eq.~\eqref{eq:rad sol} can be expressed as 
\begin{equation}
	\tilde{F}_{\text{rad}}^\mu(x) =  -2e\,\text{Re}\int\sd\Phi(q)\,iJ^\mu(q)\,e^{iq\cdot x} \;.
\end{equation}
We can express $J^\mu(q)$ in spinor form as
\begin{flalign}
	J_{\A\B} =&\ J^\mu(q)\sigma_{\mu (\A\B)} = \delta(q\cdot u)\big[mu_{(\A\B)}+i\,u_{(\A\gamma}q^{\gamma}_{\;\;\B)}\big] \nn\\[0.5em] =&\ \frac{1}{2m}\delta(q\cdot u)\big[\langle q|u|q\rangle|\bar{q}\rangle_{(\A}|\bar{q}\rangle_{\B)} - 2(q\cdot u)|q\rangle_{(\A}|\bar{q}\rangle_{\B)}\big] = \frac{1}{4m}\delta(q\cdot u)\langle q|u|q\rangle|\bar{q}\rangle_{(\A}|\bar{q}\rangle_{\B)} \;,
\end{flalign}
where we have made use of the following Schouten identities:
\begin{flalign}
	u_{(\A\B)} =&\ \frac{1}{8m^2}\big[\langle\bar{q}|u|\bar{q}\rangle|q\rangle_{(\A}|q\rangle_{\B)} + \langle q|u|q\rangle|\bar{q}\rangle_{(\A}|\bar{q}\rangle_{\B)} - 2(q\cdot u)|q\rangle_{(\A}|\bar{q}\rangle_{\B)}\big] \;,\\[0.5em] u_{(\A\gamma}q^{\gamma}_{\;\;\B)} =&\ -\frac{i}{8m} \big[\langle\bar{q}|u|\bar{q}\rangle|q\rangle_{(\A}|q\rangle_{\B)} - \langle q|u|q\rangle|\bar{q}\rangle_{(\A}|\bar{q}\rangle_{\B)} + 2(q\cdot u)|q\rangle_{(\A}|\bar{q}\rangle_{\B)}\big] \;.
\end{flalign}
And as such, the corresponding Maxwell spinor is 
\begin{flalign}
	\varphi_{\A\B}(x) =&\ \tilde{F}_{\text{rad}}^\mu(x)\sigma_{\mu (\A\B)} = -\frac{e}{m}\,\text{Re}\int\sd\Phi(q)\,i\delta(q\cdot u)\langle q|u|q\rangle|\bar{q}\rangle_{\A}|\bar{q}\rangle_{\B}\,e^{iq\cdot x} \nn\\[0.5em] =&\ \frac{e}{m}\,\text{Re}\int\sd\Phi(q)\,i\delta(q\cdot u)\langle \bar{q}|u|\bar{q}\rangle|q\rangle_{\A}|q\rangle_{\B}\,e^{-iq\cdot x} \;.
\end{flalign}
This is related to the amplitude formulation by noting that the static anyon amplitude is related via $\braket{\bar{q}|u|\bar{q}} = \frac{m}{e M}\cl{A}^{(3)}_+$.

\subsection{Gravitational Curvature Spinor}
For the gravity case, we can follow in direct analogy to the gauge theory construction. In this case, we consider the (de Donder gauge) field equation for the topologically massive graviton field $h_{\mu\nu}(x)$ in the presence of a static gravitational source $T_{\mu\nu}=Mu_\mu u_\nu\,\delta^{(2)}(\mathbf{x})$:
\begin{equation}\label{eq:TM graviton eq}
	\left(\pd^2 - m^2\right)\pd^2 h_{\mu\nu}(x) = -\mathcal{T}_{\mu\nu}(x) \;,
\end{equation}
where the effective energy-momentum tensor $\mathcal{T}_{\mu\nu}(x)$ is given by
\begin{equation}
	\mathcal{T}_{\mu\nu}(x) = 2\kappa m^2\Big[T_{\mu\nu}(x) - \eta_{\mu\nu}T(x) - \frac{1}{2m}\pd_\rho\varepsilon^{\rho\lambda}_{\;\;\;(\mu}T_{\nu)\lambda}(x) + \frac{1}{2m^2}\left(\eta_{\mu\nu}\pd^2 + \pd_\mu\pd_\nu\right)T(x)\Big] \;.
\end{equation}
We can recast \eqref{eq:TM graviton eq} into a field equation for the Cotton tensor:
\begin{equation}
	(\pd^2 - m^2)C_{\mu\nu}(x) = -\frac{\kappa}{4}\pd_\rho\varepsilon^{\rho\sigma}_{\;\;\;(\mu}\mathcal{T}_{\nu)\sigma}(x) \;,
\end{equation}
where $C_{\mu\nu}  = -\frac{\kappa}{4}\,\pd_\lambda \varepsilon^{\lambda\sigma}_{\;\;\;(\mu}\pd^2 h_{\nu)\sigma}$ in the de Donder gauge.

Given this, we can then define the radiative part of the Cotton tensor as
\begin{flalign}
	C^{\text{rad}}_{\mu\nu}(x) =&\ -\frac{\kappa}{4}\int\sd^3y\,G_{\text{rad}}(x-y)\,\pd_\rho\varepsilon^{\rho\sigma}_{\;\;\;(\mu}\mathcal{T}_{\nu)\sigma}(x)\nn\\[0.5em] =&\ -\frac{\kappa}{4}\int\sd\Phi(q)\,q_\rho\varepsilon^{\rho\sigma}_{\;\;\;(\mu}\big[\mathcal{T}_{\nu)\sigma}(q)\,e^{iq\cdot x} + \mathcal{T}_{\nu)\sigma}(-q)\,e^{-iq\cdot x}\big] \;.
\end{flalign}
In momentum space, we have that $T_{\mu\nu}(q)= M\,\delta(q\cdot u)u_\mu u_\nu$, and so the effective energy-momentum tensor is
\begin{equation}
	\mathcal{T}_{\mu\nu}(q) = 2\kappa m^2M\,\delta(q\cdot u)\Big[u_\mu u_\nu + \eta_{\mu\nu} + \frac{i}{2m}\varepsilon(q,u)_{(\mu}u_{\nu)} + \frac{1}{2m^2}\left(q^2\eta_{\mu\nu} + q_\mu q_\nu\right)\Big] \;,
\end{equation}
such that
\begin{flalign}\label{eq:cotton energy momentum}
	q_\rho\varepsilon^{\rho\sigma}_{\;\;\;(\mu}\mathcal{T}_{\nu)\sigma}(q) =&\ 2\kappa m^2M\,\delta(q\cdot u)\,q_\rho\varepsilon^{\rho\sigma}_{\;\;\;(\mu}\Big[u_{\nu)}u_{\sigma} + \frac{i}{2m}\left(\varepsilon(q,u)_{\nu)}u_{\sigma} +u_{\nu)}\varepsilon(q,u)_{\sigma}\right)\Big] \;.
\end{flalign}
As is evident from this equation, we can neglect any terms proportional to $\eta_{\mu\nu}$ or $q_{\mu}q_{\nu}$ since they will vanish in the Cotton tensor. Thus, retaining only the terms that survive upon contraction with $q_{\rho}\varepsilon^{\rho\mu\nu}$, the spinorial form of $\mathcal{T}_{\mu\nu}(q)$ is
\begin{flalign}
	\mathcal{T}_{\A\B\gamma\delta}(q) =&\ \frac{1}{4}\mathcal{T}_{\mu\nu}(q)\sigma^\mu_{(\A\B)}\sigma^\nu_{(\gamma\delta)} \nn\\[0.5em] =&\ -i\kappa mM\,\delta(q\cdot u)\big[\big(imu_{(\A\B)} - q_{(\A\epsilon}u^{\epsilon}_{\;\;\B)}\big)u_{(\gamma\delta)} + \big(imu_{(\gamma\delta)} - q_{(\gamma\epsilon}u^{\epsilon}_{\;\;\delta)}\big)u_{(\A\B)}\big] \nn\\[0.5em] =&\ -\frac{i\kappa M}{4}\,\delta(q\cdot u)\,\langle\bar{q}|u|\bar{q}\rangle\big[|q\rangle_{(\A}|q\rangle_{\B)}u_{(\gamma\delta)} +  |q\rangle_{(\gamma}|q\rangle_{\delta)}u_{(\A\B)}\big]  \;.
\end{flalign}
Given this, we can construct the spinorial form of \eqref{eq:cotton energy momentum}, the momentum space Cotton spinor
\begin{flalign}
	\Phi_{\A\B\gamma\delta}(q) =&\ \frac{1}{4}q_\rho\varepsilon^{\rho\sigma}_{\;\;\;(\mu}\mathcal{T}_{\nu)\sigma}(q)\sigma^{\mu}_{(\A\B)}\sigma^{\nu}_{(\gamma\delta)} + (\A\leftrightarrow\gamma, \B\leftrightarrow\delta)\nn\\[0.5em] =&\  \frac{1}{2}\big[q_{(\A\epsilon}\mathcal{T}^{\epsilon}_{\;\;\B)\gamma\delta} + q_{(\gamma\epsilon}\mathcal{T}^{\epsilon}_{\;\;\delta)\A\B}\big] + (\A\leftrightarrow\gamma, \B\leftrightarrow\delta)\nn\\[0.5em] =&\ -\frac{\kappa M}{4m}\,\delta(q\cdot u)\big[\langle\bar{q}|u|\bar{q}\rangle^2 |q\rangle_{\A}|q\rangle_{\B}|q\rangle_{\gamma}|q\rangle_{\delta} \nn\\ &\qquad\qquad + \langle\bar{q}|u|\bar{q}\rangle \langle q|u|q\rangle\big(|q\rangle_{\A}|q\rangle_{\B}|\bar{q}\rangle_{\gamma}|\bar{q}\rangle_{\delta} - |q\rangle_{\gamma}|q\rangle_{\delta}|\bar{q}\rangle_{\A}|\bar{q}\rangle_{\B}\big)\big] + (\A\leftrightarrow\gamma, \B\leftrightarrow\delta) \nn\\[0.5em] =&\ -\frac{\kappa M}{2m}\,\delta(q\cdot u)\,\langle\bar{q}|u|\bar{q}\rangle^2 |q\rangle_{\A}|q\rangle_{\B}|q\rangle_{\gamma}|q\rangle_{\delta} \;.
\end{flalign}
It follows that the position space Cotton spinor $\Phi_{\A\B\gamma\delta}(x)$ can be expressed as
\begin{flalign}
	\Phi_{\A\B\gamma\delta}(x) =&\ \frac{1}{4}C^{\text{rad}}_{\mu\nu}(x)\sigma^{\mu}_{(\A\B)}\sigma^{\nu}_{(\gamma\delta)} = -\frac{\kappa}{2}\int\sd\Phi(q)\,\big[\Phi_{\A\B\gamma\delta}(q)\,e^{iq\cdot x} + (\Phi_{\A\B\gamma\delta}(q))^\ast\,e^{-iq\cdot x}\big] \nn\\[0.5em] =&\ -\kappa\,\text{Re}\int\sd\Phi(q)\,\Phi_{\A\B\gamma\delta}(q)\,e^{iq\cdot x} \nn\\[0.5em] =&\ \frac{\kappa^2M}{4m}\,\text{Re}\int\sd\Phi(q)\,\delta(q\cdot u)\,\langle\bar{q}|u|\bar{q}\rangle^2 |q\rangle_{\A}|q\rangle_{\B}|q\rangle_{\gamma}|q\rangle_{\delta}\,e^{iq\cdot x} \;,
\end{flalign}
which precisely matches the amplitude calculation in eq. \eqref{cotton_anyon}.
\subsection{Reality and Dimensional Reduction Revisited}
In section \ref{motivation} we alluded to the fact that dimensional reduction can be more complicated for symmetric spinors with many indices, such as the Weyl spinor. However, scattering amplitudes expressed in spinor helicity notation are functions of momentum spinors only, which can be simply reduced using the operator $\chi^{\A\dA}$ defined in \eqref{dimRedOp}. The general prescription for converting massless spinor helicity variables in 4D to their massive equivalents in 3D is simply
\begin{equation}\label{key}
	\lambda^{\text{3D}}_\A = \lambda^{\text{4D}}_\A\bigg|_{p_2 = m},~~~~~\bar{\lambda}^{\text{3D}}_\A = \tilde{\lambda}^{\text{4D}}_{\dA}\chi^{\dA}_{~\A}\bigg|_{p_2 = m},
\end{equation}
or, in terms of the spinor products,
\begin{equation}\label{key}
	\braket{ij}_{3D} = \braket{ij}_{4D}\bigg|_{p_2 \rightarrow m},~~~~~\braket{\bar{i}\bar{j}}_{3D} = [ij]_{4D}\bigg|_{p_2 \rightarrow m}.
\end{equation}
However, we note that there is a significant problem with either $\Phi_{\A\B}$ or $\Psi_{\A\B\G\D}$ expressed in their current forms above: they vanish for real momentum. This is precisely the same issue as in four dimensional Minkowski space, and the reason is the same - there is no way to solve the on-shell condition if the momentum is real. If we consider $u^\mu = (1,0,0,0)$ in four dimensional Minkowski space, then the on-shell condition together with the energy-conserving delta function gives
\[
\delta(q\cdot u)\delta(q^2) ~~~\implies~~~ q_1^2 +q_2^2 + q_3^2 = 0.
\]
For real momentum, the only solution is $q_1^2 = q_2^2 = q_3^2 = 0$. The solution to this problem in four dimensions was to go to split signature, where the on-shell condition can be solved for real momentum as $q_2^2 = q_1^2 + q_3^2$. In three dimensions, there is no concept of split signature, however for massive particles there is another trick we can use: we can analytically continue the mass to be imaginary. In three dimensions, the on-shell condition is
\[
\delta(q_0) \delta(q^2 + m^2) ~~~\implies~~~q_1^2 + q_2^2 = -m^2. 
\]
This can be solved either for complex (or imaginary) momentum \textit{or} for $m \in i\R$. Recall that we can dimensionally reduce along the imaginary direction in spin-space, the same direction that we might choose to analytically continue to arrive at $(2,2)$. The interpretation for this imaginary mass is then clear: it is the result of a dimensional reduction from $(2,2)$ signature rather than $(1,3)$. In spinor form, this simply means taking 
\[
	\braket{ij}_{3D} = \braket{ij}_{4D}\bigg|_{p_2 \rightarrow -im},~~~~~\braket{\bar{i}\bar{j}}_{3D} = [ij]_{4D}\bigg|_{p_2 \rightarrow -im},
\]
where the four-dimensional spinors are massless Minkowski space spinors. If we are already working with $(2,2)$ spinors, then we can simply taking $p_2 = -m$ to get the same result. Overall this has a similar effect as in four dimensional split signature: the spinors and barred spinors (and polarization vectors) are completely real, now related by $m \rightarrow -m$ rather than complex conjugation. For real spinors, the polarization vectors are
\begin{equation}\label{polvectors}
	\epsilon_-^\mu(q) = -\frac{\braket{q|\gamma^\mu|q}}{2m},\quad \epsilon_+^\mu(q) = \frac{\braket{\bar{q}|\gamma^\mu|\bar{q}}}{2m}\quad.
\end{equation}

With this form of the dimensional reduction in mind, we can now relate our findings to four dimensions easily. In four dimensions with split signature, the negative helicity momentum space Maxwell spinor is given by \cite{Monteiro:2020plf}
\[
\braket{\phi_{\A\B}(q)} = -\sqrt{2}|q\rangle_{\alpha}|q\rangle_{\beta} \mathcal{A}_{+}^{(3)}(q).
\]
Let's consider the amplitude for a charged scalar, which is given by
\[
\cl{A}_+^{(3)}(q) = -2e p\cdot\epsilon_+(q) = -\sqrt{2}e\frac{[q|p\ket{\eta}}{\braket{q\eta}}.
\]
Under a dimensional reduction using $\chi^{\A\dA}\chi^{\dB}_\A = \epsilon^{\dA\dB}$ along with $\chi|q] = \ket{\bar{q}}$ and $\chi^{\dA}_{(\B}p_{\A)\dA} = 2p_{\A\B}$, this becomes 
\[\label{4d3pt}
\cl{A}_+^{(3)}(q) = -2e p\cdot\epsilon_+(q) = -\sqrt{2}e\frac{\braket{\bar{q}|p|\eta}}{\braket{q\eta}}.
\]

Choosing $\eta = \bar{k}$, we find
\[
x = \frac{\braket{\bar{q}|u|\bar{q}}}{\braket{q\bar{q}}} = \frac{\braket{\bar{q}|u|\bar{q}}}{2m}.
\]
This is the conventional definition of $x$ in three dimensions \cite{Moynihan:2020ejh,Burger:2021wss} (defined similarly to its four-dimensional cousin \cite{Arkani-Hamed:2017jhn}). We can use this to write \eqref{4d3pt} as
\[
\cl{A}_+^{(3)}(k) = -\sqrt{2}eM x,
\]
where everything is now understood as being three-dimensional and gauge-invariant, with $p_3 = -m$.  
Under dimensional reduction then, the momentum space Maxwell spinor becomes
\[
\sqrt{2}|q\rangle_{\alpha}|q\rangle_{\beta} \mathcal{A}_{+}^{(3)}(q) ~~~\longrightarrow~~~~\frac{e}{m} |q\rangle_{\alpha}|q\rangle_{\beta} \braket{\bar{q}|u|\bar{q}}
\]
This matches the expression derived directly in three dimensions, e.g. eq. \eqref{anyonspinor}. Since the momentum-space Weyl spinor is formulated in exactly the same way, it too has a simple formulation under dimensional reduction, being given analogously.

In four dimensions, the on-shell three-particle amplitudes involving massive particles was given an interpretation as computing the Newman-Penrose scalars in $(2,2)$ signature \cite{Monteiro:2020plf}. In 2+1 dimensions, three-particle amplitudes involving anyons have the interpretation of computing the Newman-Penrose scalar for a topologically massive boson with imaginary mass.
\section{Discussion}
We have established the idea of the Cotton double copy, which relates curvature spinors in both topologically massive gauge theory and topologically massive gravity. Rather pleasingly, we have shown that the vacuum expectation value of the curvature spinors in both theories can be expressed in terms of on-shell three-particle amplitudes involving a charged (massive) particle emitting a topologically massive gauge boson or graviton, where the gauge boson has imaginary mass. This follows in direct analogy to the four-dimensional double copy, as explored in \cite{Monteiro:2020plf}, and indeed, we have shown that these quantities can easily be derived via a dimensional reduction from four dimensions, specifically by taking the spinor-helicity form of Coulomb or Schwarzchild three-particle amplitudes and setting a spacetime direction to be constant with imaginary mass. 

Since the three-particle amplitudes double copy, the momentum space curvature spinors inherit this double copy, and the (dual) Maxwell spinor double copies to the Cotton spinor. We also provided an example of this in position space -- many other examples of the position space Cotton double copy, including in curved spacetimes, can be found in a paper to be released concurrently by M. Carrillo-Gonz\'alez, A. Momeni and J. Rumbutis \cite{cottondcwaves}. While we have only worked perturbatively in this paper, it is highly likely that a non-perturbative coherent state construction is plausible, as was found in \cite{Monteiro:2020plf}. 

It is clear that like its four-dimensional counterpart, the Cotton double copy presents a compelling avenue of research. In future studies, it would be interesting to further explore its implications in position space --- especially for type D spacetimes, which ought to be captured by the amplitudes above. One interesting idea is to formulate three-dimensional electromagnetic and gravitational shockwave solutions in terms of amplitudes, in analogy to the four-dimensional set-up recently explored in \cite{Cristofoli:2020hnk}, as these have also been shown to obey the Cotton double copy \cite{cottondcwaves}.

\textit{Note}: During the course of this project we became aware of the paper by M. Carrillo-Gonz\'alez, A. Momeni and J. Rumbutis \cite{cottondcwaves}, which has some overlap with our work. We thank them for sharing their results with us and for the useful correspondence.

\section*{Acknowledgements}
NM would like to thank Mariana Carrillo Gonz\'alez, Tristan McLoughlin, Donal O'Connell, Matteo Sergola and Guy Jehu for useful discussions. NM is supported by STFC grant ST/P0000630/1 and the Royal Society of Edinburgh Saltire Early Career Fellowship. WTE is supported by the Czech Science Foundation GA\v{C}R, project 20-16531Y.
\appendix
\section{Conventions}
We choose a \textit{real} $SL(2,\R)$ spinor basis of three-dimensional spacetime that satisfies the Clifford algebra\footnote{We define $$A^{(\mu} B^{\nu)} \equiv A^\mu B^\nu - A^\nu B^\mu$$}
\begin{equation}\label{key}
	{\gamma^{(\mu}\gamma^{\nu)}}^{\A}_{~\B} = 2\eta^{\mu\nu}\delta^\A_{~\B},
\end{equation} 
where
\begin{equation}\label{paulibasis}
	\sigma^\mu_{\A\B} = \left\{\begin{pmatrix}
		1 & 0 \\
		0 & 1
	\end{pmatrix},\begin{pmatrix}
	0 & 1 \\
	1 & 0
\end{pmatrix},\begin{pmatrix}
1 & 0 \\
0 & -1
\end{pmatrix}\right\},~~~~~\epsilon_{\A\B} = \begin{pmatrix}
0 & -1 \\
1 & 0
\end{pmatrix} = -\epsilon^{\A\B}
\end{equation}
\begin{equation}\label{key}
	\left(\gamma^{\mu}\right)^\A_{~\B} = \epsilon^{\A\G}\sigma^\mu_{\G\B} = \left\{\begin{pmatrix}
		0 & 1 \\
		-1 & 0
	\end{pmatrix},\begin{pmatrix}
		1 & 0 \\
		0 & -1
	\end{pmatrix},-\begin{pmatrix}
		0 & 1 \\
		1 & 0
	\end{pmatrix}\right\}.
\end{equation}
We note that the $SL(2,\R)$ basis of Pauli matrices is symmetric. Throughout the text, we make repeated use of the following identities
\[\label{epsID}
\varepsilon^{\mu\nu\rho}\sigma_{\rho\,\A\B} = \eta^{\mu\nu}\epsilon_{\A\B} + (\sigma^\mu\sigma^\nu)_{\A\B}
\]
\[
\sigma^{\mu}_{\A\B}\,\sigma_{\mu\,\G\D} = 2\varepsilon_{\A\D}\varepsilon_{\B\G} + \varepsilon_{\A\B}\varepsilon_{\G\D}=-\varepsilon_{\A(\D}\varepsilon_{\G)\B}
\]

Angle brackers in 2+1 dimensions are given in terms of spinors as
\begin{equation}\label{key}
	\ket{i}_{\A} = \lambda_{i\A},~~~~~\bra{i}^{\A} =  \lambda_{i}^\A,~~~~~	\ket{\bar{i}}_{\A} = \bar{\lambda}_{i\A},~~~~~\bra{\bar{i}}^{\A} =  \bar{\lambda}_{i}^\A.
\end{equation}
Spinor inner products are given by
\begin{equation}\label{key}
	\braket{ij} \equiv \epsilon^{\A\B}\lambda_{i\B}\lambda_{j\A} = \lambda_{i}^{\A}\lambda_{j\A} = - \lambda_{i\A}\lambda_{j}^{\A}.
\end{equation}
Fourier transforms are defined via
\[
	&f(x)=\int \hat{\sd}^{4} q e^{-i q \cdot x} \tilde{f}(q), \\
&\tilde{f}(q)=\int \sd^{4} x e^{i q\cdot x} f(x) .
\]
\bibliographystyle{JHEP}
\bibliography{cottondc} 
\end{document}